%% file: main.tex
\documentclass[lettersize,journal]{IEEEtran}

\usepackage{cite}
\usepackage{amsmath,amssymb,amsfonts}
\usepackage{algorithmic}
\usepackage{graphicx}
\usepackage{textcomp}
\usepackage{xcolor}
\usepackage{gensymb}
\usepackage{url}
\usepackage{adjustbox}
\usepackage{multirow}
\usepackage{nicematrix}
\usepackage{mwe}
\usepackage{caption}
\usepackage{subcaption}
\usepackage{multirow}
\usepackage{colortbl}
\ifCLASSINFOpdf

\begin{document}

\title{S-RAN: Semantic-Aware Radio Access Networks}

\author
{Yao Sun,~\IEEEmembership{Senior Member,~IEEE}, Lan Zhang,~\IEEEmembership{Member,~IEEE}, Linke Guo,~\IEEEmembership{Member,~IEEE}, Jian Li,~\IEEEmembership{Member,~IEEE}, Dusit Niyato,~\IEEEmembership{Fellow,~IEEE}, and Yuguang Fang,~\IEEEmembership{Fellow,~IEEE}
       
\thanks{Yao Sun is with the James Watt School of Engineering, University of Glasgow, Glasgow G12 8QQ, U.K. (email: Yao.Sun@glasgow.ac.uk).

Lan Zhang (corresponding author) and Linke Guo are with the Department of Electrical and Computer Engineering, Clemson University,  South Carolina, USA (e-mail: \{lan7; linkeg\}@clemson.edu).

Jian Li is with the School of Cyber Science and Technology, University of Science and Technology of China, Hefei, China (e-mail: lijian9@ustc.edu.cn).

Dusit Niyato is with the School of Computer Science and Engineering, Nanyang Technological University, Singapore (e-mail: dniyato@ntu.edu.sg)

Yuguang Fang is with the Department of Computer Science, City University of Hong Kong, Hong Kong SAR, China (e-mail: my.fang@cityu.edu.hk).
}
}

\maketitle

\begin{abstract}
Semantic communication (SemCom) has been a transformative paradigm, emphasizing the precise exchange of meaningful information over traditional bit-level transmissions. However, existing SemCom research, primarily centered on simplified scenarios like single-pair transmissions with direct wireless links, faces significant challenges when applied to real-world radio access networks (RANs). This article introduces a Semantic-aware Radio Access Network (S-RAN), offering a holistic systematic view of SemCom beyond single-pair transmissions. We begin by outlining the S-RAN architecture, introducing new physical components and logical functions along with key design challenges. We then present transceiver design for end-to-end transmission to overcome conventional SemCom transceiver limitations, including static channel conditions, oversimplified background knowledge models, and hardware constraints. Later, we delve into the discussion on radio resource management for multiple users, covering semantic channel modeling, performance metrics, resource management algorithms, and a case study, to elaborate distinctions from resource management for legacy RANs. Finally, we highlight open research challenges and potential solutions. The objective of this article is to serve as a basis for advancing SemCom research into practical wireless systems.
\end{abstract}

\IEEEpeerreviewmaketitle

\section{Introduction}\label{Introduction}

\input{introduction}

\section{S-RAN Architecture}

\input{architecture}

\section{S-RAN Transceiver Design} 
\input{link}
\section{S-RAN Radio Resource Management}
\input{network}

\section{Open Research Topics}
\input{OpenTopics}

\section{Conclusion}
This article has presented a new paradigm of semantic-aware RAN (named S-RAN) to enable novel wireless communication schemes for semantic communications. We have first outlined the S-RAN architecture, and then illustrated the design principles and challenges on both end-to-end transmission and multi-user networking, along with some initial results. Finally, we have highlighted several open topics associated with S-RAN. It is believed that this work stands as a pioneer in exploring the realm of futuristic semantic communications, offering a comprehensive perspective from the entire RAN.

\ifCLASSOPTIONcaptionsoff
  \newpage
\fi

\bibliographystyle{IEEEtran}
\bibliography{reference.bib}







\end{document}

%% file: introduction.tex
Semantic communication (SemCom) has recently emerged as a transformative paradigm, prioritizing accurate reception of message meaning over mere bits~\cite{chaccour2024less,luo2022semantic,lu2022rethinking}. The core concept is to empower wireless transceivers with reasoning and causality. In a typical SemCom, the source transceiver first employs a semantic encoder to refine semantic features and filter out extraneous content, thereby reducing the required number of bits while preserving the original message semantic meaning. Subsequently, the destination transceiver utilizes a semantic decoder to precisely recover source information, even in the presence of significant bit errors at the syntactic level. Importantly, by integrating background knowledge related to source messages between transceivers, SemCom holds immense potential for efficient exchanges of desired information with low semantic ambiguity and fewer bits.

While SemCom shows great promise, it is still in its infancy, attracting research interest from various perspectives. One fundamental research area involves the development of rigorous theoretical modeling to comprehensively and mathematically characterize this novel transmission paradigm. This includes quantifying semantic information, measuring channel capacity with the gain of semantic reasoning, and evaluating information distortion at a semantic level~\cite{xie2021deep,yan2022resource,lu2022rethinking}. Another popular research avenue focuses on designing optimal transceivers tailored to specific tasks, such as image, text, and audio, with the aim of reducing transmission data volume while maintaining high-quality perception~\cite{luo2022semantic,xie2021deep}. Additionally, new transmission protocols have been proposed to enhance performance for both a single and multiple transmission pairs, 
introducing new metrics from a semantic perspective~\cite{yan2022resource,xia2023joint,xia2023wiservr}.

Despite considerable progress in SemCom,  most existing literature like \cite{luo2022semantic,lu2022rethinking,xie2021deep} primarily assumes a simple-pair transmission scenario with a direct wireless link between transceivers. This assumption neglects the complexities of real-world radio access networks (RANs). In a well-established cellular system, for instance, terminal devices (TDs) are typically associated with a nearby base station (BS), establishing wireless uplinks and downlinks for sending and receiving data through RANs. 
Several unique challenges arise when implementing SemCom in RANs and endowing RANs with reasoning ability and causality.
\begin{itemize}
    \item \textit{RAN Architecture Evolution}: Legacy RANs are primarily designed based on bit-oriented transmission, without taking into account of knowledge base (KB) and semantic coding. To overcome this limitation, the RAN architecture must evolve to accommodate SemCom requirements by incorporating new physical entities and logical functions to perform semantic coding and KB embedding.
    \item \textit{Transceiver Adaptation}: Existing SemCom transceivers are trained in an end-to-end (E2E) manner with specific channel conditions, rendering them unsuitable for direct use in RANs with complicated multi-hop channel conditions and diverse hardware platforms.
    \item \textit{Protocol Development}: It is crucial to systematically discuss and develop new protocols covering the PHY layer and MAC layer within RANs. Particular attention should be directed towards multi-user access with limited radio resources, as this plays a pivotal role in providing dedicated SemCom gain from the system perspective.
    \item \textit{Knowledge Management}: SemCom's unique demands for background knowledge pose significant challenges in constructing and updating this KB within a single RAN and across multiple RANs. Facilitating infrastructure to harness dynamic and personalized knowledge to achieve high semantic accuracy and generality is challenging.
\end{itemize}

To tackle these challenges, this article develops a \underline{S}emantic-aware \underline{R}adio \underline{A}ccess \underline{N}etwork (S-RAN) framework, aiming to explore SemCom from a holistic RAN perspective rather than solely focusing on a single transmission pair. 
We start with the introduction to the architecture of S-RAN, detailing both its physical and logical components while discussing key challenges based on case studies. Subsequently, we delve into the design of S-RAN transceivers, taking into account crucial features such as distinct uplink and downlink channel conditions, limited background knowledge, and asymmetric reasoning capabilities. Additionally, we investigate S-RAN radio resource management, underscoring significant distinctions and design principles, along with empirical studies. Finally, open research challenges and potential solutions are discussed to fully unlock S-RAN capabilities.

%% file: architecture.tex
\begin{figure*}
  \includegraphics[width=\textwidth]{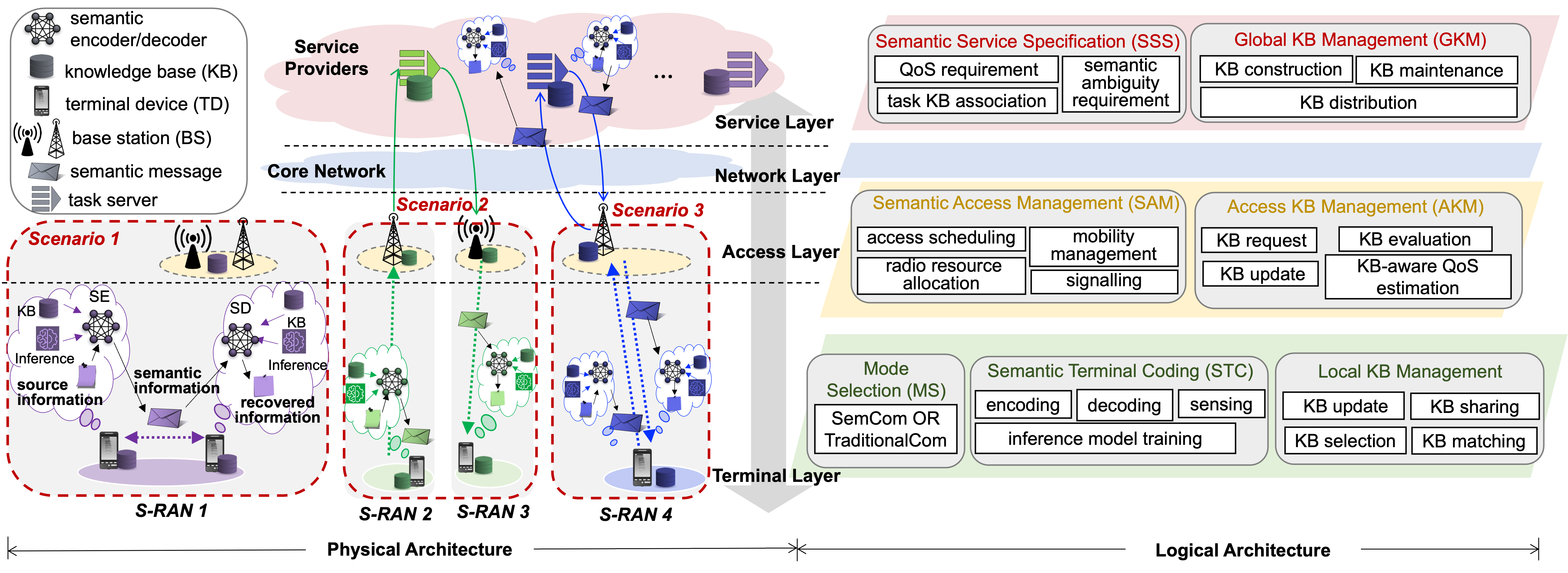}
  \caption{S-RAN Architecture Overview. The physical architecture (\textit{Left}) includes legacy physical infrastructures with additional new entities, where three typical access scenarios are illustrated. The logical architecture (\textit{Right}) involves four key layers, including the terminal, access, network, and service layers, in order to accommodate SemCom under the physical architecture.} 
  \label{fig:arch}
\end{figure*}

S-RAN aims to offer a holistic solution to underpin SemCom. 
Despite substantial efforts, legacy RANs are primarily designed based on Shannon's classic information theory to optimize bit-level system utilities. This poses significant challenges to fulfilling the objectives of S-RAN, which emphasize semantic reasoning and causality. 
This section introduces S-RAN architecture by involving additional physical entities and logic protocols while retaining all other functionalities of the legacy RAN architecture.

\subsection{Physical Architecture}\label{sec:phy}
As illustrated in Fig. \ref{fig:arch}, wireless systems typically consist of four main layers, with the RAN consisting of the bottom two layers: the terminal layer and the access layer. 
The terminal layer consists of TDs, equipped with semantic encoders and decoders for extracting and interpreting semantic information. While most of the physical entities in S-RAN perform the same functionalities as legacy wireless systems, KB as the key element in S-RAN should be elaborated. There are several representation formats for KB, including knowledge graphs, semantic labels, and factual triples~\cite{liang2024generative}. Note that S-RAN does not specify a particular format for KBs, and both the format and content of KBs can vary significantly for different services. For each service, service providers (i.e., servers) located in the service layer are responsible for constructing and updating global KBs, which contain general information commonly used. In the access layer, we assume that wireless infrastructures, say BSs in this work, periodically distribute relevant global KBs to their associated TDs. Additionally, BSs periodically collect information from TDs and upload it to service providers to help global KB management. In the terminal layer, TDs store portions of global KBs along with personal knowledge/information/preferences, forming their personalized KBs for encoding and decoding. Wireless infrastructures in the access layer can provide supplementary computing resources and KB storage to support battery-powered TDs in processing SemCom tasks.

We discuss three typical communication scenarios in S-RAN. Note that all three scenarios could coexist in practice, forming a hybrid access environment. 
\begin{itemize}
    \item \textit{Scenario 1: terminal-to-terminal SemCom within the same S-RAN.} Most existing SemCom research focuses on this scenario. Two proximal TDs can directly transfer semantic information through a direct wireless link. The semantic encoder and decoder can be trained based on the KB stored locally at TDs or remotely at the BSs and task servers. No semantic information traffic will be passed through the BS, but the BS coordinates radio control functions. 
    \item \textit{Scenario 2: terminal-to-terminal SemCom across different S-RANs.} The source and destination TDs in different S-RANs engage in SemCom through the core network. 
    As shown in Fig.\ref{fig:arch}, the source TD within S-RAN 2 first employs a semantic encoder to generate the semantic message, which is then uploaded to the associated BS, traversing the core network to reach the task server. The task server assesses this request and forwards the messages to the destination TD in S-RAN 3 for decoding. 
    The management of KBs, encompassing their construction, updating, and sharing, require alignment with service specifications, communication link quality, available radio resources, etc. \cite{liang2024generative}. 
    \item \textit{Scenario 3: terminal-server SemCom through the S-RAN.} A TD connects to a task server for SemCom, which is typical for TDs to acquire remote services, such as uploading or downloading content on social media platforms. In this scenario, the task server functions as a terminal, equipped with the semantic encoder and decoder to infer and interpret semantic information. The task-specific KB can be deployed at the task server or the BS for terminal-side encoder and decoder management. 
\end{itemize}

\begin{table*}[!thb]
\renewcommand\arraystretch{1.0}
\caption{S-RAN Design Challenges: scenario-specific challenges (access scenarios in Fig.~\ref{fig:arch}) and common challenges.}\label{Table1}
\resizebox{\linewidth}{!}{
\begin{tabular}{clc|cc}
\hline
\rowcolor[HTML]{EFEFEF} 
 & \multicolumn{2}{c|}{\cellcolor[HTML]{EFEFEF}S-RAN Transceiver Design}  & \multicolumn{2}{c}{\cellcolor[HTML]{EFEFEF}S-RAN Radio Resource Management} \\ \hline
\rowcolor[HTML]{EFEFEF} & \multicolumn{1}{c|}{Scenario-specific Challenge} & Common Challenge & \multicolumn{1}{c|}{Scenario-specific Challenge} & Common Challenge \\ \hline
\cellcolor[HTML]{FFFFC7} &\cellcolor[HTML]{FFFFC7}&\cellcolor[HTML]{EFEFEF}&\cellcolor[HTML]{FFFFC7}&\cellcolor[HTML]{EFEFEF} \\ 
{\multirow{-2}{*}{\cellcolor[HTML]{FFFFC7}Scenario 1}}& {\multirow{-2}{*}{\cellcolor[HTML]{FFFFC7}\begin{tabular}[c]{@{}c@{}} transceiver coordination\\(without BS supervision) 
\end{tabular}}} & \cellcolor[HTML]{EFEFEF} &{\multirow{-2}{*}{\cellcolor[HTML]{FFFFC7}\begin{tabular}[c|]{@{}c@{}}
resource competition \\with cellular systems\end{tabular}}} & \cellcolor[HTML]{EFEFEF} \\ \cline{1-2} \cline{4-4}
\cellcolor[HTML]{FCFF2F} &\cellcolor[HTML]{FCFF2F}&\cellcolor[HTML]{EFEFEF}&\cellcolor[HTML]{FCFF2F}&\cellcolor[HTML]{EFEFEF} \\ 
{\multirow{-2}{*}{\cellcolor[HTML]{FCFF2F}Scenario 2}}  & {\multirow{-2}{*}{\cellcolor[HTML]{FCFF2F}\begin{tabular}[|c|]{@{}c@{}}channel asymmetry\\(uplink vs. downlink)\end{tabular}}} & \cellcolor[HTML]{EFEFEF} &{\multirow{-2}{*}{\cellcolor[HTML]{FCFF2F}\begin{tabular}[c|]{@{}c@{}}resource mismatch\\ (S-RANs 2 vs. 3)\end{tabular}}} & \cellcolor[HTML]{EFEFEF} \\ \cline{1-2} \cline{4-4}
\cellcolor[HTML]{FFC702} &\cellcolor[HTML]{FFC702}&\cellcolor[HTML]{EFEFEF}&\cellcolor[HTML]{FFC702}&\cellcolor[HTML]{EFEFEF} \\ 
{\multirow{-2}{*}{\cellcolor[HTML]{FFC702}Scenario 3}}    & {\multirow{-2}{*}{\cellcolor[HTML]{FFC702}\begin{tabular}[|c|]{@{}c@{}}capacity mismatch\\(computing and storage)\end{tabular}}}
& \multirow{-6}{*}{\cellcolor[HTML]{EFEFEF}\begin{tabular}[c]{@{}l@{}}- complicated channel conditions\\ - background knowledge embedding, \\ \ \ updating, and alignment \\ - heterogeneous hardware platforms\end{tabular}} & {\multirow{-2}{*}{\cellcolor[HTML]{FFC702}\begin{tabular}[|c|]{@{}c@{}}resource bottleneck\\(fronthaul vs. backhaul)\end{tabular}}} & \multirow{-6}{*}{\cellcolor[HTML]{EFEFEF}\begin{tabular}[c]{@{}l@{}}- semantic channel modeling\\ - system-level evaluation metrics\\ - unique knowledge management \\ - computing and communication \\ \ \ trade-off\end{tabular}} \\ \hline
\end{tabular}}
\end{table*}


\subsection{Logic Functions and Protocols}

Generally, while S-RAN does not require new interfaces for bit-level data exchanges, protocol modifications to semantically guide bit-level data exchanges are necessary in two aspects: updating existing function modules, and introducing new function modules as shown in Fig.\ref{fig:arch}(right).

Most of the existing function modules remain unchanged, while particular modules such as cell selection/association, CSI detection, radio resource allocation, power control, need to be optimized accounting to SemCom features. Specially, we will discuss the CSI detection required in S-RAN transceiver design in Section III, and resource management especially cell selection and bandwidth allocation in Section IV.

Besides, new modules are required in S-RAN. This magazine article will not discuss how to optimize each new module but explain the function achieved by each module. Specifically, the terminal layer involves three new modules. Mode selection is required to determine whether SemCom or traditional communication scheme should be selected once a call is arrived. This decision making is tied to RAN status (such as available resources, traffic load, KBs, etc.) and service requirements. The local KB management module maintains the updated terminal KB and aligns terminal KBs between source and destination, enabling efficient SemCom within S-RAN. The semantic terminal coding module handles semantic encoding and decoding while managing dynamic channel conditions and various background knowledge conditions through the implementation of sensing and inference model training functions.  
The access layer introduces two new logic modules. Similarly, the access KB management module oversees background knowledge for all terminals within S-RAN. Specifically, the KB request and update functions ensure the S-RAN with the latest KBs from task servers, while the KB-aware QoS estimation function evaluates a TD's semantic access performance based on its KB. Additionally, the semantic access management module oversees terminal access and radio resource allocation by considering semantic-level system utility. Furthermore, in the service layer, two new logic modules are incorporated. The global KB management module manages KB construction, maintenance, and distribution from the perspective of the service provider. Meanwhile, the semantic service specification module enforces task-specific semantic service performance, guiding the measurement of semantic utility within S-RANs and managing KBs throughout the entire system.

\begin{figure}[!tb]
\centerline{\includegraphics[width=0.5\textwidth]{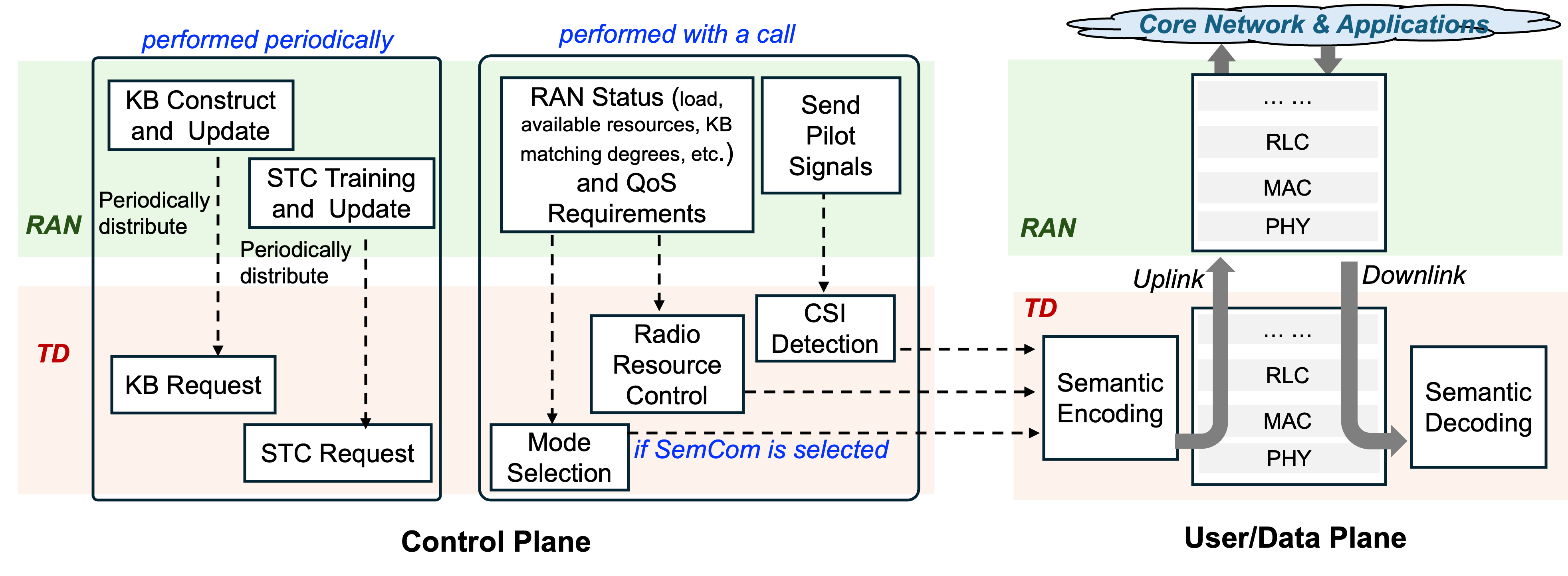}}
\caption {An illustration of data flow in S-RAN.}
\label{fig:dataflow}
\end{figure}

Fig.\ref{fig:dataflow} presents the data transmission flow in practice. Note that only the unique parts in S-RAN are presented, while most of the other unchanged modules are omitted for simplicity. Generally, the protocol is divided into the control plane and the data/user plane. As usual, the control plane is responsible for control signaling exchanges, where KB management and semantic terminal coding (STC) management are performed periodically to support SemCom encoding and decoding at TDs. For the data/user plane in S-RAN, taking the uplink as an example, when a call arrives, the TD needs to make/obtain the mode selection decision (SemCom or traditional communication) according to radio resource availability, traffic load, KB matching degrees, etc. If the traditional communication scheme is selected, all the protocols remain the same for the data transmission. If SemCom is selected, new radio resource allocation schemes as discussed in Section IV will be applied. Given the allocated radio resources, the CSI will be detected in a traditional way using pilot signals, and the CSI information alongside the allocated resources will be provided into a semantic encoding module to semantically convert source data into binary bits for transmission. The data should pass by PHY, MAC, RLC, etc. protocol stakes, which are with necessary modifications as discussed.

\subsection{S-RAN Design Challenges}
Table~\ref{Table1} outlines potential design challenges based on the above discussions, focusing on transceiver design and radio resource management from the perspectives of a single SemCom transmission and multiple transmission pairs. Regarding transceiver design, common challenges encompass complicated channel estimation, KB embedding and evolution, and hardware heterogeneity, while common challenges in radio resource management involve semantic channel modeling, system-level utility evaluation, unique knowledge management, and the trade-off between computing and communication. Additionally, to draw attention to various and intricate S-RAN scenarios, specific issues for different transmission scenarios in Fig.~\ref{fig:arch} are presented. Taking Scenario 1 as an example, without BS coordination, TDs encounter challenges related to coordination and radio access competition—competing with those directly connected to BSs. In Scenario 2, TDs accessing different BSs from distinct S-RANs may face mismatched radio resources due to independent radio management in S-RANs. In Scenario 3, TDs may lack the capability to match the semantic interpretation capability at the task server. All aforementioned challenges highlight the broad scope of investigation and optimization within the realm of S-RAN. 

%% file: link.tex

\begin{figure*}
  \center\includegraphics[width=0.9\textwidth]{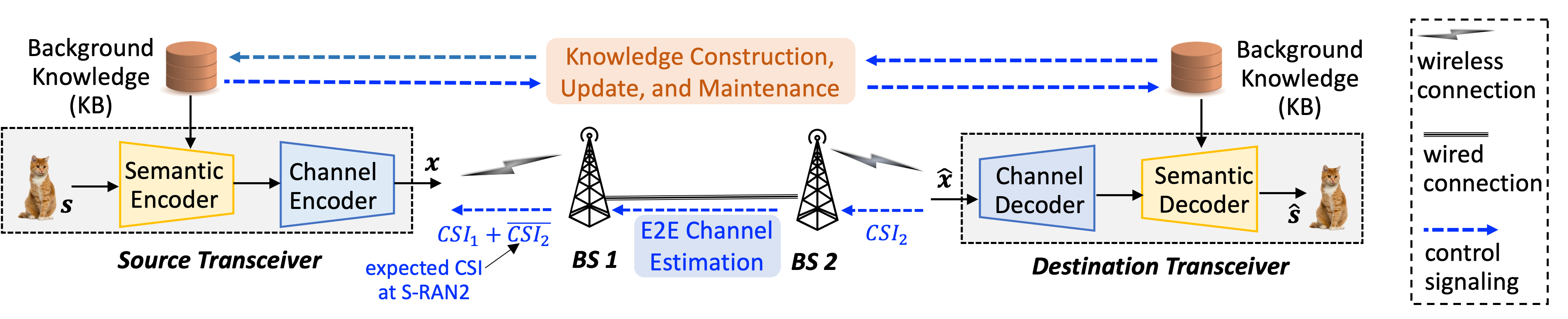}
\caption {S-RAN transceiver design entails wireless infrastructures, such as BSs, collaborating with transceivers to facilitate the end-to-end channel estimation and knowledge management.
}\label{fig:transceiver}
\end{figure*}


Fig. 2 shows the framework of the S-RAN transceiver design, where KBs are equipped with two terminals. A semantic encoder and channel encoder are deployed at the transmitter to generate bits $x$ being transmitted from the source message $s$, while a semantic decoder and channel decoder are deployed at the receiver to recover the message $\hat{s}$ from channel-distorted bits $\hat{x}$.
Although lots of efforts have been put in SemCom transceiver design, the challenges below still remain. 
\begin{itemize}
    \item \textit{Static channel conditions}. Conventional training assumes available instantaneous channel state information (CSI), such as certain noise or fading conditions. However, in situations like Scenario 2 in Fig.\ref{fig:arch}, where transceivers are associated with different BSs, this assumption becomes unrealistic in most cases. 
        
    \item \textit{Oversimplified background knowledge models}. The assumption that source and destination transceivers share the same KB might not hold true due to the distributed nature of transceivers \cite{zhang2023prompt}. This lack of consistent KBs across SemCom tasks can result in insufficient semantic extraction and interpretation for joint source and channel coding at a semantic level (denoted as S-JSCC).
    
    \item \textit{Resource-constrained hardware platforms}. Transceiver platforms are typically assumed to possess sufficient storage and processing ability to perform S-JSCC based on DNNs. However, many transceiver platforms are battery-powered and resource-constrained, presenting challenges in managing the costs associated with computing, storage, and power consumption of DNNs. 

\end{itemize}

\subsection{Augmenting Channel Awareness}
In contrast to bit-level communications, the quality of SemCom relies significantly on S-JSCC at both the source and destination TDs. The E2E channel information plays a key role in both the training and inference process of the S-JSCC against channel variations upon the observed channel conditions. As intermediate nodes, BSs lack the ability to scrutinize or enhance transmission quality at the semantic level, therefore, transceivers must perceive the E2E channel conditions before sending or receiving SemCom messages. 
As shown in Fig.\ref{fig:transceiver}, in addition to capturing the CSI of the link between the source and BS1, it is necessary to obtain the CSI of the link between the destination and BS2. This CSI estimation is incorporated into the SAM module of the access layer in Fig.\ref{fig:arch} with the cooperation of BSs and TDs.
This augmentation for channel awareness necessitates a careful trade-off between signaling traffic, computational cost, and the effectiveness of channel estimation.

\subsection{Enforcing Background Knowledge Alignment}
To enhance reasoning and causality abilities, it is crucial for both the source and destination transceivers to share the same KB for proper extraction and accurate interpretation of semantic information. In S-RAN, where task-specific KBs vary, we focus on efficient KB management across TDs. However, distributing KBs to mobile TDs is challenging due to their varying preferences, access association, and local resource budgets. To address this, we introduce KB management modules for the terminal, access, and service layers as shown in Fig.\ref{fig:arch}, facilitating swift KB alignment while avoiding unnecessary costs for the network system and TDs. Specifically, the updated KB at the service layer will be distributed to the access layer based on the task popularity for each S-RAN and the update magnitude of KBs. Once received, the access layer will convey the update notice in control signaling. When engaging in SemCom, source and destination TDs will first exchange local KB version numbers, and then the TD with lower version number will download the new version to ensure the consistency of KB at both ends. KB management scheme is needed to decide whether and when to update local KBs by considering signalling and computing costs. 



\subsection{Accommodating Hardware Platform Heterogeneity}
Given resource-constrained transceiver platforms, the cost of storing and processing S-JSCC built upon DNNs becomes a major concern. Recent research has introduced hardware-efficient ML techniques~\cite{deng2020model}, shedding light on efficient S-JSCC design. Neural network prunings~\cite{deng2020model}, for example, can significantly reduce the size of DNNs by searching and removing unimportant parameters. To balance efficiency and effectiveness, the pruning process is usually computationally intensive. Therefore, in S-RAN, pruning can be executed on powerful infrastructures, such as BSs, to adaptively construct hardware-efficient S-JSCC, aligning with different transceiver platforms. This functionality can be incorporated into the SAM module of Fig.\ref{fig:arch}. An alternative approach is to enable transceivers to offload S-JSCC tasks to BSs depending on TDs' resource budgets. 

%% file: network.tex
 Following the discussion of the S-RAN transceiver design, the focus now shifts to examining the resource management among transceivers at the access layer. Below key aspects need to be examined when performing radio resource management.

\subsection{Semantic Channel Modelling} \label{Subsec_Channel}

The first challenge of resource management in S-RAN is to develop a mathematically sound channel model that aligns well with the salient characteristics of SemCom. 
In contrast to the traditional information theory, which relies on stochastic correlation of source information, the SemCom system demands the measurement of semantic information with full consideration of logical connections among semantics. Importantly, a key feature of S-RAN is the deployment of KB for information reasoning, which introduces a critical distinction in quantifying the amount of semantic information, as interpretation can vary under different KBs.
For example, a TD with specific background knowledge that associates entity A (a particular desk) with entity B (a printer on that desk) could obtain extra information from entity A. Therefore, quantifying semantic information should carefully consider 1) the logic connection among items in the information and 2) the impact of KBs. 

Quantifying semantic information can be approached from various angles. A common method involves adapting classical information concepts, such as entropy, while considering logic connections and KBs to measure the amount of information at a semantic level \cite{carnap1952outline, bao2011towards, choi2022unified}. 
Logic probability, especially when formulated with a specific KB, can effectively capture reasoning ability \cite{bao2011towards, choi2022unified}. Another perspective is to utilize category theory \cite{bradley2022enriched}, particularly in language/text transmission, to mathematically describe the logic relation (so called morphism in category theory) among semantics (modeled as elements in category theory), thereby approximating the meaning of source information (an expression abstracted in category theory).

Once a method for quantifying semantic information is established, the subsequent step involves 
modeling the semantic capacity of a wireless channel. This entails considering two crucial factors: 1) the physical channel capacity for transmitting bits, 2) the reasoning ability of the communication pair under a given physical channel. 
For the first factor, we have traditional Shannon capacity to measure the maximum number of bits to be transmitted under a given channel condition. The second factor unique to S-RAN, reasoning ability, relates to the KB matching degree between the pair as well as the coding model capability (both encoder and decoder). One common way to model KB is using graph theory, where the matching degree can be derived based on the structure of KB graphs. For coding model capability, it is possible to adopt the theory of deep learning to estimate 
the accuracy and generalization ability of the models, thus to mathematically model reasoning ability for a given encoder-decoder pair with KBs under given channel conditions. In an ideal scenario, assuming the same encoder and decoder models are used for all communication pairs, the coding capability remains consistent \cite{xia2023joint}. Once quantifying semantic information and channel capacity, other indicators such as semantic distortion/ambiguity and semantic data rate can be derived.  

\subsection{Performance Metrics}

Traditional metrics designed for measuring bit-level communication performance are no longer suitable when assessing S-RAN. In response to this shift, we introduce dedicated performance metrics crafted specifically for S-RAN. 

\begin{itemize}
    \item \textit{Link-level performance metrics.} Unlike the bit/symbol error rate, the focus in SemCom shifts to semantic ambiguity or semantic similarity, reflecting the accuracy of recovered meaning. Semantic ambiguity/similarity depends on knowledge matching degree between the transmission pair, encoder and decoder capabilities, and the physical channel condition. For text transmission, measuring word differences between the source and recovered text using Bilingual Evaluation Understudy (BLEU) \cite{xie2021deep} and calculating sentence similarity \cite{yan2022resource} are common approaches to get semantic ambiguity/similarity. In addition, image and video tasks may employ metrics such as Peak Signal-to-Noise Ratio (PSNR) \cite{xia2023wiservr}.
    \item \textit{System-level performance metrics.} Expanding the scope to S-RAN with multiple transmission pairs involves studying performance metrics at the system level. A key metric is message throughput, defined as the number of messages correctly recovered within the network \cite{xia2023joint}. Calculating message throughput in S-RAN necessitates employing advanced channel modeling, as discussed in Section \ref{Subsec_Channel}, taking into account factors such as physical channel conditions, coding capability, and background knowledge. Additionally, semantic spectrum efficiency, proposed in \cite{yan2022resource}, measures transmission efficiency at a semantic level, indicating the amount of semantic information received per second per Hertz. Furthermore, a related metric for assessing transmission efficiency in S-RAN from an energy perspective, termed semantic energy efficiency, can be defined.
\end{itemize}

\subsection{Radio Resource Management Algorithm Design Principles}

Once suitable metrics for assessing S-RAN performance are identified, the subsequent focus shifts to designing algorithms for optimizing radio resource allocation. 
It is worth mentioning that while the solution to radio resource management can still be based on traditional mathematical tools, such as optimization theory, game theory, reinforcement learning, etc., the challenge lies in how to formulate the radio resource allocation problem to intrinsically capture S-RAN features. Consequently, when formulating and designing resource management in S-RAN, the following distinctive factors should be considered.

\begin{itemize}
    \item \textit{KB matching degree:} In S-RAN, where source and destination TDs may possess different KBs, the degree of KB matching between the pair plays a crucial role in minimizing semantic ambiguity \cite{xia2023joint}. 
    Consequently, when formulating resource management problems in S-RAN, accounting for KB matching degrees becomes paramount. Decisions regarding resource allocation should weigh factors such as higher KB matching degree with worse physical channel condition versus lower matching degree with better physical channel condition, aiming to enhance overall system performance.
    \item \textit{Coding ability of TDs:} 
    S-RAN employs semantic and channel coding modules based on diverse DNNs, reflecting different computing resources of TDs. This diversity in coding abilities can lead to significant variations in transmission efficiency. Hence, resource management algorithms need to factor in the distinct coding abilities of TDs.
    \item \textit{Channel condition:} Clearly, channel conditions should be a pivotal consideration in the design of resource allocation algorithms, given their influence on both physical channel capacity and the performance of the encoder and decoder in S-RAN. The channel modeling discussed earlier needs to be incorporated into the formulation of resource allocation problems. 

    \item New performance metrics: As discussed in Section IV.B, new performance metrics at the semantic level should be introduced to examine S-RAN performance at the semantic level. These new metrics could complicate the resource allocation problem, as metrics like message throughput might exhibit randomness due to the uncertainty in KBs, semantic coding models, semantic ambiguity, etc. Thus, these new metrics could make resource optimization problems stochastic and less tractable.
    
    \item \textit{Multiple S-RAN Scenarios:} 
    In S-RAN, where all three scenarios coexist shown in Fig.\ref{fig:arch}, resource management complexities arise. The implications of resource allocation policies extend to both intra-scenario TDs and inter-scenario TDs. For instance, a downlink TD in scenario 3 may generate interference for users in scenarios 1 and 2, leading to a degradation of SemCom performance. Therefore, the design of resource management policies should carefully consider the performance of both intra-scenario and inter-scenario TDs.
\end{itemize}

\subsection{Case Study}
Similar to that in Fig. 3, we consider a use case in S-RAN consisting of multiple semantic TDs and BSs, where imperfect knowledge matching (IKM) between TDs and BSs is assumed. As previously discussed, S-RAN encompasses three distinct scenarios, and in this study, we specifically focus on TDs from scenarios 2 and 3, i.e., we do not include D2D communications in our case. We aim to develop a resource management algorithm to determine user association and bandwidth allocation as per the design principles described above. Specifically, we choose system throughput in message (STM) \cite{xia2023joint} as the metric rather than traditional bit-level metrics to measure the number of messages that can be corrected and recovered at destination TDs in the S-RAN network. The constraints on resource budget and KB matching requirements are considered. The two intuitive baselines based on physical channel conditions without considering SemCom features are exploited (one is the max-SINR plus water-ﬁlling algorithm, and the other is the max-SINR plus evenly distributed algorithm) to verify the performance gain of the proposed algorithm.


\begin{figure}[!tb] 
\centerline{\includegraphics[width=0.38\textwidth]{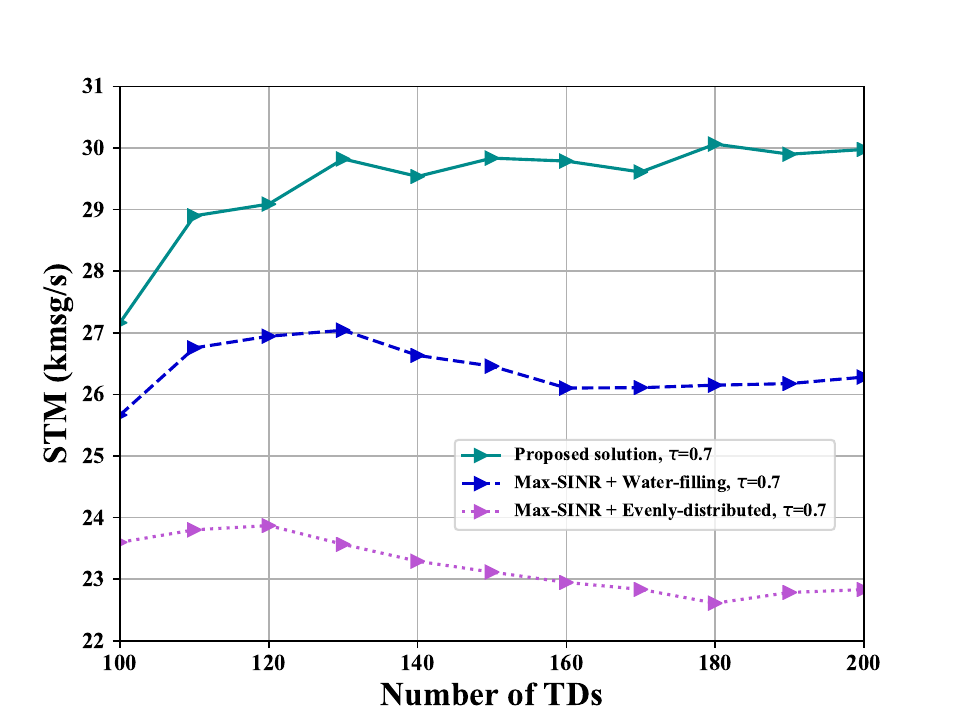}}
\caption{Performance of system message throughout under different number of TDs. 
}\label{fig:sim1}
\end{figure}

\begin{figure}[!tb]
\centerline{\includegraphics[width=0.38\textwidth]{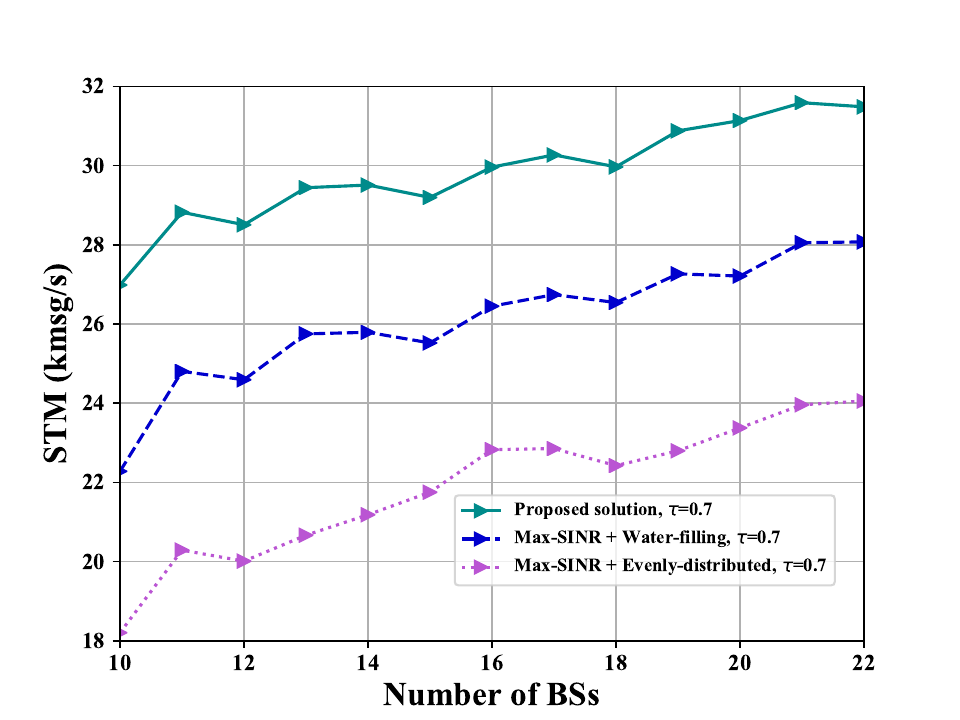}}
\caption{Performance of system message throughout under different number of BSs. 
}\label{fig:sim2}
\end{figure}

Fig. \ref{fig:sim1} ﬁrst shows the comparisons of STM with varying numbers of TDs under a mean knowledge matching degree $\tau$ of 0.7. The number of BSs is fixed as 16. It can be seen that our solution can always gain an extra STM performance around 2 kmsg/s to 6 kmsg/s with the increasing number of TDs when compared with the two baselines.  Fig. \ref{fig:sim2} shows the similar performance gain under 200 TDs with varying numbers of BSs.  Both simulations validate the effectiveness of our radio resource management principles, which emphasize the consideration of SemCom relevant factors in addition to physical channel conditions.

%% file: OpenTopics.tex
Despite its anticipated superiority, the practical deployment of S-RAN faces several challenges, giving rise to the following open research topics. 

\subsection{Lack of Theoretical Framework}
A fundamental challenge lies in the absence of rigorous theoretical modeling to mathematically measure the entire SemCom process, covering aspects such as semantic information quantification, semantic channel capacity, and information distortion at a semantic level. Shannon's classical theory, developed for conventional bit-level communications using pre-defined codebooks, 
fall short when applied to SemCom. This limitation arises because these theories view data correlation solely from a stochastic perspective, overlooking the logical connections among semantic elements. Consequently, accurately modeling and evaluating the performance of S-RAN becomes challenging without a matched theoretical framework, hindering optimization of network performance. 

\subsection{Hybrid Data Transmission Schemes}

Basically, with the coexistence of SemCom and conventional bit-level communication users in future wireless networks, managing such a hybrid network presents unique challenges to synergistically improve resource utilization. For instance, traditional communications prioritize bit-related performance, while SemCom emphasizes semantic-level performance. How to unify performance metrics for this hybrid network should be the first difficulty. Then, in terms of resource algorithm design, how to accommodate extra features of SemCom such as KBs, semantic ambiguity, and semantic encoders/decoders while subject to the constraints of conventional communications should be another non-trivial point. In addition, multi-hop data transmission with SemCom and bit-level communication could be another challenging issue in this hybrid network. The performance of each hop might be evaluated using different metrics, and the latency of each hop can fluctuate significantly depending on the data volume being transmitted and encoding/decoding schemes. Consequently, optimizing the end-to-end transmission across multi-hops becomes even more challenging.

\subsection{Network Dynamics} 
Transceivers in S-RAN are commonly trained under specific channel conditions/distributions and fixed knowledge. However, the network dynamics, characterized by varying channel conditions, fluctuating available resources, incremental knowledge, and diverse task requirements, can significantly influence both the semantic accuracy of end-to-end data transmission and the overall network-level performance. Consequently, designing transceivers with strong generalizability and robustness becomes a noteworthy challenge. Although some efforts to exploit advanced machine learning algorithms show promise, deploying S-RAN in practical scenarios remains challenging due to the highly complicated process of SemCom-enabled data transmission.